# IOT BASED SMEs SHOP MANAGEMENT SYSTEM

UMMAE HAMIDA HANNAN
Faculty of Science and Technology Department of Computer Science
American International University-Bangladesh shagufa.hannan@gmail.com

MD REAZ UDDIN CHOWDHURY
Faculty of Science and Technology Department of Computer Science
American International University-Bangladesh reaz000723.rc@gmail.com

MD. GOALM RAHAMAN
Faculty of Science and Technology Department of Computer Science
American International University-Bangladesh golamrahaman0@gmail.com

SAKIB MAHMUD GALIB
Faculty of Science and Technology Department of Computer Science
American International University-Bangladesh sakib7355@gmail.com

DR. MD. TAIMUR AHAD
Assistant Professor Faculty of Science and Technology Department of Computer Science
American International University-Bangladesh taimur.ahad@aiub.edu

**Abstract—**The Internet of Things (IoT) is an idea that intends to interface arranged data frameworks to actual items. The Internet of Things (IoT) has applications in pretty much every part of life in this day and age, and stock administration is no special case. IoT gives an answer for this issue by making it simpler to interface every one of the various organizations in a strategic framework utilizing Wireless Sensor Networks. An Interactive Shopping Model and an Automated Inventory Intelligent Management System that uses the Internet of Things to give constant item following, the board, and observing. A study and examination of the commonness of IoT among assembling SMEs is introduced, just as the current impediments and possibilities for permitting prescient investigation. The four examination capacities are depicted alongside an outline of the IoT empowering agents. Future patterns and difficulties in arising innovative work subjects are featured, for example, making IoT advances available to SMEs. The motivation behind this paper is to look at how the Internet of Things is changing our lives and work spaces, just as to feature probably the best strategic approaches, insights, and patterns. Considering the developing significance of big business IoT and the exploration hole in this field, an IoT design and the IoT administration industry will be examined. A model is needed to choose and send IoT administrations in different authoritative settings.

**INTRODUCTION**

The present innovation is progressing dangerously fast, particularly in developed nations. Everybody needs everything to be simpler and quicker. The Internet of Things (IoT) is an organization that empowers actual gadgets, well-being, administrations, business, vehicles, security, and different things inserted with programming, hardware, actuators, sensors, and organization network to gather and trade information. Since the Internet of Things was made to further develop productivity, diminish human intercession, and increment precision, the article can be detected or controlled remotely utilizing the Internet. The client can screen the framework in a good way utilizing the IoT strategy, and there is no requirement for direct contact with the framework. —AC, Ceiling Fan, Window, Smoke Detector, Light, Siren, Fire Sprinkler, Thermostat, Power Meter, Battery, Door (Shutter), Motion Detector, Webcam (v CCTV), Car (Smoke Generator), Heating Element (Fire), Solar Panel, PC, Router, Printer, Switch, Smartphone. Everything can be utilized remotely with an assortment of cell phone applications. This is because of the framework's productivity and new innovation. Besides, in light of its low power utilization and usability, the remote controller by telephone has forever been a choice. A shop chief, for instance, can utilize IoT to control their security framework. The objective of the venture is to help little retailers in keeping up with the security framework while they are away from the store. The framework can be utilized in any telephone application to guarantee that a shop representative can arrive at the shop from any area. We can likewise check out their brilliant shopping basket charging framework as a principal part. The remote sensor is associated with the forced air system, roof fan, and window, which likewise has a LCD screen, an IR sensor.The remote association will be associated with each thing accessible in the shop. The Internet of Things has as of late acquired broad acknowledgment in an assortment of fields. As per numerous specialists, the Internet of Things will totally change how individuals communicate with their environmental factors, bringing about the development of a billion- dollar industry that will be the new main thrust for data innovation extension. The web of things alludes to the most common way of associating objects to the web and permitting them to impart and trade data with each other to accomplish savvy ID, continuous following, observing, and the board. Radio recurrence ID (RFID), electronic name (EPC), remote detecting innovation, worldwide situating framework, and standardized identification per user are only a couple of the data detecting advances utilized in the Internet of Things. The Internet of Things is an idea that alludes to a data sharing biological system in which heterogeneous gadgets (things) are associated through wired or remote organizations. The Internet of Things (IoT) is an improvement of machine-to-machine (M2M) correspondence that involves the coordination of sensors, actuators, and other inserted gadgets by means of an IP-based systems administration model. IoT empowers the utilization of Wireless Sensor Networks (WSN) to gather information from an assortment of sensors, which is then traded and envisioned in reality with the assistance of actuators. The Internet of Things (IoT) is comprised of countless entertainers, including sensors, things, sensor organizations, actuators, and people, who all add to its appropriate work. With such progressions in innovation, IoT related to Cloud processing can be utilized to make a mechanized stock administration framework, bringing about the advancement of a brilliant shopping complex. Many stock administration frameworks have been proposed throughout the most recent few decades, yet they needed elements like continuous checking, detect ability, on-the-fly stock information refreshing, e-installment, information examination, and secure client validation. Conventional frameworks for counting and overseeing co ordinations depended on manual techniques or standardized tag innovation, yet they couldn't stay aware of the developing number of clients and enhanced co-ordinations.

The Internet of Things, when utilized related to a stock administration framework, can support the improvement and the board of all coordination-related exercises, accordingly expanding productivity and consumer loyalty. The reason for this paper is to introduce an Automated Inventory Management System (AIMS) for further developed coordination and a shrewd shopping complex. The fourth modern unrest incorporates the modern web of things (IoT) and computerized reasoning (AI). Little and medium-sized endeavors (SMEs) come up short on the assets and information to utilize Industry 4.0's high level and present-day innovations and techniques.

**LITERATURE REVIEW**

Parts of the Internet of Things Environment Technological The Physical Setting Hardware Various remote gadgets are utilized to associate human and non-human items to the Internet of Things, permitting correspondence and connection between them over a universal remote organization. [4] There are not many examinations on the IoT biological system and engineering that are pertinent to the development of big business IoT. For example, we actually don't completely grasp what undertaking design components are and how they help in the improvement of explicit endeavor

IoT administrations. The undertaking IoT environment depicted in the past area gives the innovation stages needed to IoT design execution. While scientists still can't seem to settle on an IoT engineering, they typically utilize a complex methodology, with each layer devoted to explicit capacities like correspondence/detecting, information handling, and information handling/thinking. [5]

Coordinated stages that permit the IoT's different equipment, programming, and systems administration components to cooperate consistently. Norms Various specialized and functional principles characterize the plan and intractability of different IoT components. [4] In request to create fruitful IoT administrations, IoT design should be set up and refreshed consistently to commission and decommission different IoT resources and administrations. Wireless organizations are more normal in IoT applications in light of the fact that numerous IoT applications require unlimited development of items inside an actual space. According to a study of SMEs, current cloud arrangements are inadmissible for them, supporting the improvement of SME-explicit cloud arrangements. Discoveries on the qualities of SMEs Manufacturing SMEs, as recently expressed, have a one-of-a-kind arrangement of attributes when contrasted with bigger organizations. Two of them zeroed in on making a learning manufacturing plant to show SMEs Industry 4.0 basics. As indicated by a writing audit, the subject of IoT in SMEs is still somewhat new, and SMEs ought to embrace new plans of action to stay cutthroat. They all concur that SMEs have potential, which legitimizes the improvement of SMEs-explicit adaptations. [2]

Coming up next are the significant advantages of the Internet of Things that affect business: Communication - The Internet of Things (IoT) permits associated gadgets and clients to keep a steady association and trade information. The Internet of Things idea, which alludes to all associated objects that trade information, is broadly utilized in all parts of life, including wearable and different gadgets and sensors that make each item shrewd. [3] In the social IoT, there are different proprietors/substances associating through their gadgets, while collective endeavor IoT applications are regularly run by a solitary administering body. A coordinated effort can occur between individuals, among individuals and things, and between things in the IoT framework. Entrepreneurs assume a basic part in the IoT's turn of events.

These business people utilize their specialized information, business experience, and instinct to make new plans of action in the IoT domain, roused by a craving for self-gain, self-realization, or local area commitment. Rejuvenating these business thoughts often requires tending to exist specialized, administrative, and legitimate issues by growing new advancements, new business processes, and coming to an obvious conclusion in various alternate ways identified with IoT. [4] The Internet of Things ought to be incorporated as per a particular vision and thought, distinguish openings for utilizing innovation, draw in business organizations and government, and encourage a culture of Internet of Things utilization. [3]

**PROPOSED WORK**

The system would allow for real-time asset monitoring, management, and end-to-end system trace ability. The system would serve as a guide, directing him to the locations of his desired products. Overall, the system would be automated, resulting in a more efficient system. This system's goal would be to create an interactive shopping environment. The system would allow for real-time asset monitoring, management, and end-to-end logistics chain trace ability. It is based on Internet of Things (IoT) wireless technology and cloud computing.

**Small and medium-sized enterprises:**

At the local, regional, national, and European levels, encouraging innovation in small and medium-sized enterprises remains at the heart of policy initiatives aimed at stimulating economic development. On a theoretical level, innovation has also displaced efficiency as the primary focus of much theory development and policy analysis, with efficiency becoming a necessary complement to innovation. Despite the increased focus on the role of SMEs in innovation, there is a gap between what is understood by the general innovation literature and what is known about innovation in SMEs. Despite the lack of a common theoretical foundation for innovation research in general, it is clear that studies of innovation in SMEs have largely failed to reflect advances in the innovation literature. This failure to improve our fundamental understanding of SMEs is disappointing, given that SMEs account for 99 percent of businesses in the UK, 55 percent of non-governmental employment, and 51 percent of turnover. [11]

**Infrastructure Layer:**

In this section, we will go over all of the actors (physical entities) that make up the system, their roles, basic functionalities, and how they communicate with one another. The workflow of our proposed system for such an interactive shopping model is depicted in the figure below.

**Sensors:**

Sensors serve as the system's eyes and ears, detecting events and environmental conditions and transmitting the information gathered. The sensors' job is to observe and perceive physical world events or phenomena. Sensors are classified into three types based on three factors: sensor type, methodology, and sensing parameters. The sensor type determines whether the sensor is homogeneous or heterogeneous, as well as whether it is single-dimensional or multidimensional.

Methodology refers to the methods used by a sensor to collect data. It can be active or passive in nature. Sensing parameters are the number of factors that a sensor can detect. A sensor can detect a single parameter, like body temperature, or multiple parameters, like an ECG.

**Gateway:**

Sensors serve as the system's eyes and ears, detecting events and environmental conditions and transmitting the information gathered. The sensors' job is to observe and perceive physical world events or phenomena. Sensors are classified into three types based on three factors: sensor type, methodology, and sensing parameters. The sensor type determines whether the sensor is homogeneous or heterogeneous, as well as whether it is single-dimensional or multidimensional. Methodology refers to the methods used by a sensor to collect data. It can be active or passive in nature. Sensing parameters are the number of factors that a sensor can detect. A sensor can detect a single parameter, like body temperature, or multiple parameters, like an ECG

**Mobile Device:**

A mobile device acts as a customer service representative. Customers' mobile devices enable them to search for and purchase items in real-time. Any notification or special offer relating to a specific product is immediately sent to the mobile device, which then awaits a response.

**RESEARCH GAP**

Data security and data protection, data quality, the usage of common standards and protocols, inseparability, legal difficulties, and so on are the key issues and challenges facing the Internet of Things, just as they are for Internet-based technologies. Other major challenges facing the Internet of Things, as outlined by (6), include establishing a common addressing mechanism for effective device addressing, developing low-cost embedded devices that are more energy- efficient and reliable, establishing governing bodies to oversee device usage, establishing quick and reliable communication, and minimizing the load on servers as well as embedded devices. Increased production and deployment of these devices go beyond resolving these concerns, and all actors, particularly businesses and governments, must work together to resolve them in a timely manner and create policies that ensure the correct and authorized use of the Internet of Things.

One of the most significant concerns is privacy, which specifies the rules under which individual data can be accessed. The identification and tracking of devices, as well as all of their behaviors and the extraction of personal data from application areas, make users feel like they are being monitored and can be tracked at any time (7). On the one hand, it is convenient to locate persons who have gotten lost, have been abducted, or have been in an accident. On the other side, it is an inconvenience for anyone who wants to maintain their privacy. One of the most major challenges that the Internet of Things must address is security. Affordable and low-cost broadband connections, as well as Wi-Fi capabilities in many gadgets, are required for simple locations in public spaces, leaving them vulnerable to cyber-attacks. With a special focus on IoT systems, the Internet of Things permits permanent data sharing between connected objects and defines three main components for maintaining security: authentication, secrecy, and access control. (8)

Compatibility is a term used to describe how well two things work together. Different manufacturers of smart device sensors and platform providers utilize different data transfer protocols, which might lead to communication issues. (9). There are numerous organizations and projects aimed at establishing communication standards.

Complexity is an important factor to consider. Complex systems that connect and manage various things are vulnerable to failure. Users should not put too much faith in technology because defective scanners and sensors can result in incorrect data or no data being sent. Electromagnetic disturbances, vibrations, and human age could all play a role. Eye and finger injuries, for example, can impact scanning the iris of an eye or a fingerprint for bio-metric data.(10).

**EXPERIMENTAL METHOD**

In today's world, supermarket shopping is becoming increasingly popular. It is now recommended that a smart shopping complex based on IoT avoid this issue. It includes an Android app, Wi-Fi, and the cloud. Customers can use a search tool to learn about the mall's product availability and unavailability. Wi-Fi, LCD, Android app, Amazon cloud, and billing unit central can all be used to wirelessly connect all types of shop equipment. The goal of this proposed system is to provide a system that can be used to address small shop management through the Internet of Things. We've added a simulation image over here, where we can see that all of the equipment is connected via a smartphone and a desktop computer, both of which are connected to the internet via the router. In the shop, the computer is connected to a switch. The smart wireless IOT device here is the home gateway, which is also connected to the switch. Our shop management system's main and core device, the home gateway, is wirelessly connected to all devices. For security purposes, our shop has two CCTV cameras, CCTV 1 and CCTV 2, both of which are open 24 hours a day and connected to the internet. We also have a motion detector that connects to those cameras to detect people motion. 2 lights – Light and Light 1, Light 1 is open 24 hours a day and is connected to the door that acts as a shutter. 2 fans, 4 windows, and an air conditioner are connected to the home gateway and have some conditions to use those devices in different temperatures to control the shop's internal environment temperature. We use a temperature monitor to check the temperature here (temperature monitor shows the shop temperature). A smoke detector, fire sprinkler, fire detector, and siren are installed in the shop for fire and safety reasons. It will notify that customer that there is an emergency in this store. A car serves as the smock generator, and a heating element serves as the fire. The solar panel is also used here to convert solar energy to electricity. The solar panel was connected to the battery and stored energy in it. This battery provides a source of power for light 1. All of those devices can be operated/controlled remotely using a smartphone or a computer.

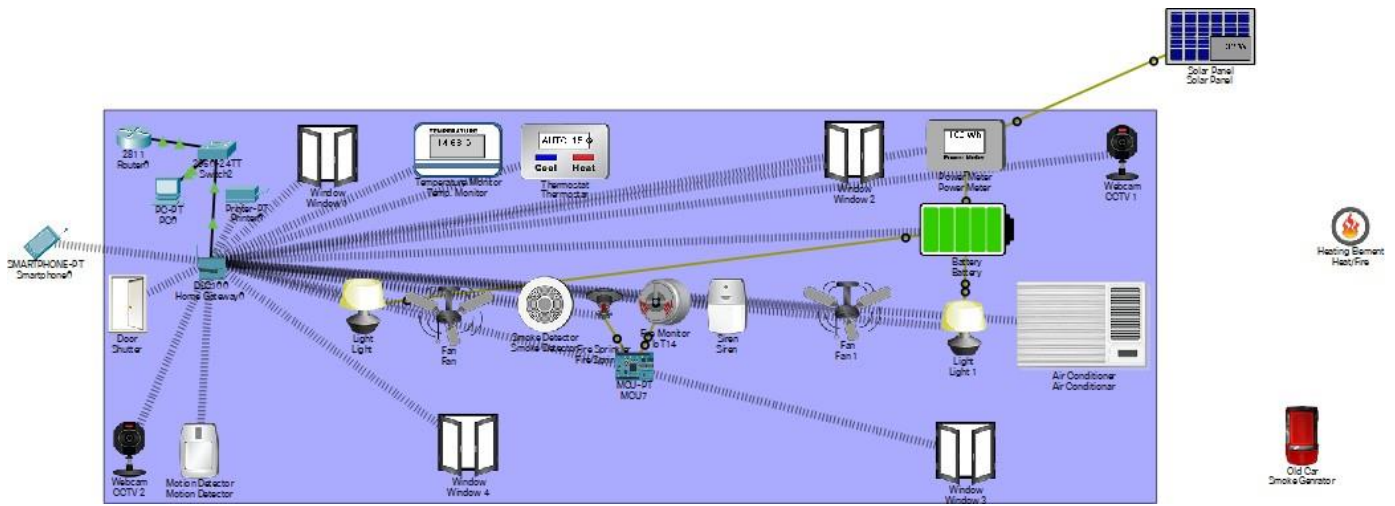

Fig. 1. IoT based SMEs shop management system simulation

**RESULT**

Nowadays, the Internet of Things (IoT) is a complete smart solution. Our lives are made easier by

IoT devices, particularly in the business and home sectors. Using IoT devices, we can control fans, air conditioning, and windows in our small shop management system by measuring temperature. If the temperature drops below 10°C, all windows, fans, and air conditioning will be turned off. Between 11°C and 15°C degrees Fahrenheit, the windows will be open. Low- temperature fans will be set between 12°C and 15°C degrees Celsius. Fans will be turned on at temperatures ranging from 15°C to 20°C degrees Celsius. If the temperature rises above 20°C, the air conditioner will automatically turn on, and the windows will close. After 22°C, the fan speed drops to a low level. If the temperature falls below 20°C, the air conditioner will turn off automatically, and the fan speed will be set to high. If the door to our shop is open and unlocked, all lights will be turned on. In this simulation, we use two lights, one of which is powered by a battery and the other by a direct power source. When electricity is unavailable, the battery will serve as a power source. Here, a small solar panel is used to charge the battery, and the power meter displays the amount of electricity produced by the solar panel. This shop has a smoke generator, fire sprinkler, and fire monitor siren for security and safety. If a smoke generator detects smoke, it will sound the alarm and open all windows. If the fire monitor detects a fire, the fire sprinkler will automatically activate, and the power to all devices will be turned off, but the windows will remain open. In general, all of our devices are intelligently connected to the door. All lights, fans, windows, air conditioning, and a printer will turn off automatically if the door is closed and locked. The CCTV camera is operational 24 hours a day, seven days a week. If any of the cameras are turned off, we can turn them on remotely. When the door is closed and locked, a motion detector helps detect any unusual activity or people. When the door is unlocked and opened, all of the store's lights will automatically turn on. All of these devices can be accessed locally and remotely via Wi-Fi from the shop computer.

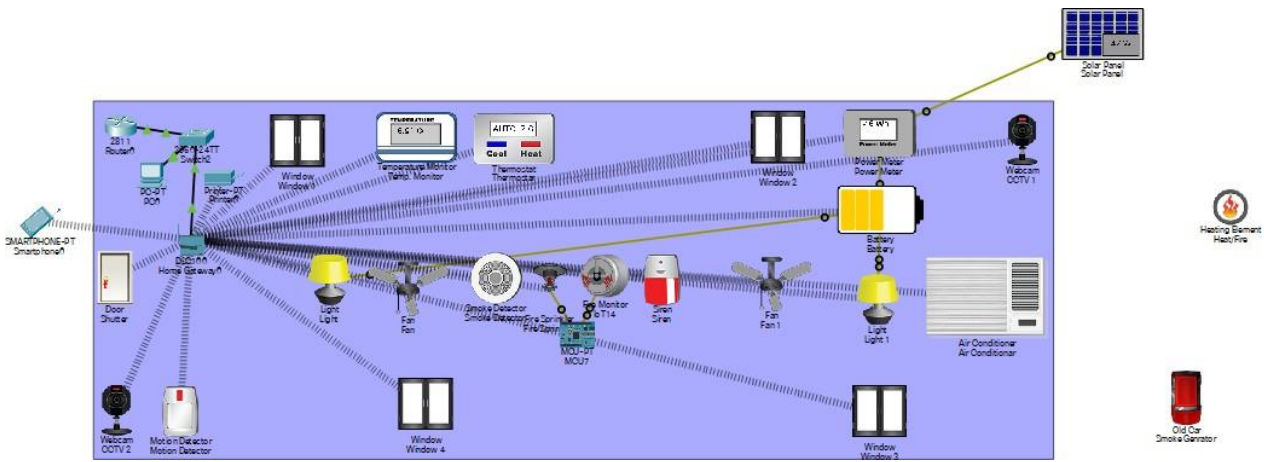

Fig. 2. Siren On when door is lock and motion detector on.

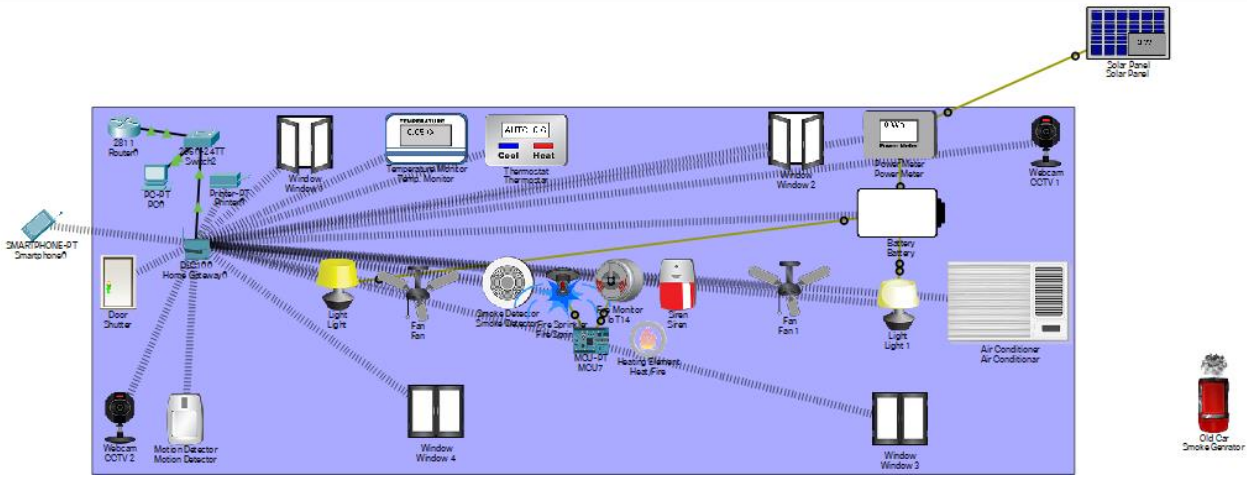

Fig. 3. Siren and fire sprinkler on when detect Smoke Fire.

**DISCUSSION**

One of the many areas where IoT can be used is in retail management. Integration of advanced or new marketing technologies, such as IoT, AI, or Bots, in the Romanian market, can be a challenge in terms of specific costs, particularly for SMEs [13]. According to studies, Romania is still not ranked highly in terms of digitization when compared to other European countries, coming in second to last after Bulgaria. Nonetheless, the country's "digital economy" [14] shows strong growth potential. Although the Internet of Things concept is not as popular among SMEs in the IT industry as it has been in other studies [15–16], it has tremendous potential because IoT adoption and integration is a top priority for the future. According to other studies [17], more than half of the analyzed companies (64.2 percent) have already integrated some of the new technologies into their marketing processes. Despite the fact that Artificial Intelligence was the most popular technology among the analyzed companies, two of them had integrated IoT technologies. Companies that do not expand their operations to new locations use advanced marketing technologies only infrequently, whereas those with more than four locations use them more frequently. A business that is able to operate in multiple locations is thought to have more financial clout. During the simulation of our proposed system, we attempted to maintain it using a smartphone with Wi-Fi that we could control remotely. The simulation's results are displayed using smartphone apps to control our devices such as fans, lights, fire monitoring, fire sprinklers, air conditioning, CC cameras, motion sensors, motion detectors, fire detectors, temperature sensors, window and door, and so on. We can manage our shop with our smartphone and ensure its security to protect it from any type of claims, such as theft, safety, and fire issues, while also conserving energy. When the shop is empty, it turns off all lights and closes the door to conserve energy.

**FUTURE WORK**

In the future, SMEs in management may be required to use robust machine learning methods to ensure better quality control and production monitoring. Machine-level predictive analytics could be enabled by smart IoT devices attached to a single machine. Data would be uploaded to a cloud service, which would store it and make it available for management reports. Small shop management is one of the IoT's smaller components, in which we attempt to demonstrate how we can manage a small shop using IoT. We only put the management system's structure in place here. We paid no attention to the security and privacy features that could be implemented in the future.

We can also add more equipment if necessary. We can create a mobile application to control the entire system as well in near future.

**CONCLUSION**

Because the Internet of Things is such a new phenomenon, there is some research on enterprise IoT. This makes it difficult for businesses to make well-informed judgments about IoT service development. To respond to the disruptive nature of IoT innovation and offer new and transformation IoT services, businesses must keep an eye out for freshly created technologies. It provides a real-time view of the entire logistics process, improving inventory efficiency by reducing operational costs and saving human resources. We discovered that low costs and simple implementation of IoT and cloud solutions were key factors in many of the successful use cases. This supports Moeuf et al. that cloud solutions are the most used in SMEs because of their simplicity in comparison to the rest of Industry 4.0. Future research should concentrate on making other aspects of Industry 4.0, such as the Internet of Things and artificial intelligence, easier to use and implement. This could be in the form of a ready-to-use product that can be implemented at any point in the manufacturing process without the need for computer scientists to set up. Choosing an IoT platform is a significant decision that necessitates effectively exploiting the Iot network for the development of many IoT enterprise apps and services. While many applications require standardised, low-performance, and low-power IoT platforms, high-performance platforms are still required to meet the computational demands of data-intensive IoT applications and edge devices. Without a doubt, the rapid development of new technologies has an impact on all aspects of daily life, and the analytic and reports presented show that these trends will continue and grow in the coming years. We are all part of this technological revolution, whether we like it or not, and the most important thing is to learn how to use it properly and wisely. The Internet of Things should be integrated according to a specific vision and idea, identify opportunities for using technology, attract business institutions and government, and create a culture of Internet of Things use. In addition to selecting the appropriate platform, businesses must decide whether they will install their own internal IoT infrastructure, third-party IoT infrastructure, or a hybrid of the two.